\documentclass[11pt,a4paper]{article}

\usepackage{subfig}
\usepackage{amsmath} 
\usepackage{graphicx}
\usepackage{authblk}

\title{Cyber-Automotive Simulation and Evaluation Platform for Vehicular Value Added Services}
\author[*]{Raja Sattiraju}
\author[*]{Pratip Chakraborty}
\author[*]{Hans D. Schotten}
\author[**]{Xiaohai Lin}
\author[**]{Daniel Goerges}
\affil[*]{Institute for Wireless Communication and Navigation, University of Kaiserslautern} 
\affil[**]{Juniorprofessorship for Electromobility, University of Kaiserslautern}

\begin{document}

\maketitle

\begin{abstract}
An easily moving and safe transportation is an indicator in any country in the world of economic growth and well-being. For the past 100 years, innovation within the automotive sector has brought major technological advances leading to safer, cleaner and more affordable vehicles. But for the most time since the inception of the moving assembly line for vehicle production by Henry Ford, the changes have been incremental / evolutionary. Thanks to the new possibilities due to the IT / wireless revolution, the automotive industry appears to be on the cusp of revolutionary change with potential to dramatically reshape not just the competitive landscape but also the way we interact with vehicles, and indeed the future design of our roads and cities. Apart from connected personal mobility, vehicles are also envisioned to provide Value Added Services (VAS) such as autonomous driving via Vehicle-to-Vehicle (V2V) and Vehicle-to-infrastructure (V2I), electric load balancing via Vehicle-to-Grid (V2G) solutions, communication solutions using Visible Light Communications (VLC) etc. The development and evaluation of vehicular VAS requires a modular and scalable multidisciplinary simulation platform. In this paper we propose a novel simulation platform named Cyber-Automotive Simulation \& Evaluation Platform (CASEP). The purpose of CASEP is to evaluate and visualize the gains of various vehicular VAS with special emphasis on commercial vehicle VAS. The use cases are evaluated with respect to the mission-specific performance indicators, thereby providing usable metrics for optimization. The visualization platform is being developed using the UNITY 3D engine, thereby enabling intuitive interaction as in real physics-based games.\

\end{abstract}

\section{Introduction}
The desire to go where we want whenever we want has been a powerful market force for centuries and the automotive industry has been and continues to be an essential component of the German economy, with a domestic turnover of 125,000 mil Euros and a foreign export value exceeding 350,000 mil Euros. The German automotive industry has been steadily raising adding 54,000 new jobs since 2010. In order to be continually sustainable, the automotive industry today is evolving from a simple manufacturer to a mobility service provider. The recent formation of Mobility as a Service (MaaS) alliance by 20 odd technological partners comprising manufacturers, universities and standardization bodies can be seen as a driving force for building up momentum to address the mobility needs of a dense urban 2020 informative society.

Direct Vehicle-to-Vehicle (V2V) and Vehicle-to-Infrastructure (V2I) communications based on co-operative awareness have proven to increase passenger safety and traffic efficiency and are currently being integrated into the production vehicles today. However, if the perception of a vehicle changes to that of a "personal mobility device" as envisioned by the MaaS alliance, novel and futuristic use cases can be constructed. Commercial vehicles such as trucks and logistic vehicles can provide Value Added Services using V2V and V2I thereby generating revenue. These include not only essential services like traffic safety and efficiency, but also services that make our lives more enjoyable such as semi/full autonomous driving, data broadcasting showers, remote vehicle control etc. Future services may include smart mobility solutions for both urban and highway scenarios where vehicular moving networks form the back of the Internet of Things (IoT). 

Commercial vehicles offer a broader scope of VAS especially with smart logistics (METIS 5G \cite{osseiran2014scenarios}) and smart public transportation systems. Both urban transportation systems and commercial trucks in highways can be used to provide novel VAS such as platooning lead services, mobile coverage extensions etc. These commercial vehicles can also subscribe to other VAS such as smart traffic signaling services that help it determine optimal speeds within a city. Public safety vehicles can also benefit from such services such as location-based broadcasting and prioritized signaling in real time etc. Factors such as degree of autonomy and degree of co-operation determine the emergence of vehicular VAS in the future. Another recent example of a use case is "Over the air" update features for electronic control units.

\subsection{UC1: Peak Power Balancing with Plug-in Hybrid Electric Vehicles (PHEVs) }
The peak electricity demand (usually between 16:00 - 17:00) can be balanced by using the energy from the PHEVs battery packs. The PHEVs parked in locations such as offices, stadiums and arenas can be used as energy containers discharging electric power during peak hours and charging during off peak hours. A suitable business model needs to be worked out that provides incentives (for example, in the form of energy credits) to the owner for offering the services of his vehicle. Charging and discharge cycles can be pre-programmed into the vehicle optimizing parameters such as time spent on parking, charge required to drive home, desired discharge cycles that can be spent by the vehicle's owner etc.). Related models, parameters, and algorithms can be developed and evaluated with a simulation and evaluation platform in a very flexible way.

\subsection{UC2: Mobile Base Stations disseminating Content on Demand}
Vehicles equipped with robust backhaul links to the 5G mobile network can be used as moving networks facilitating wireless communication between other vehicles and Road Side Units (RSUs). The information can be regarding
\begin{itemize}
	\item parking spots (with both charging/discharging options)
	\item localized radio/navigation maps
	\item context specific content broadcast
	\item mobile base stations for wireless coverage/availability enhancement
	\item co-operative safety and efficiency.
\end{itemize}
The functionality of such mobile base stations can be investigated with the simulation and evaluation platform.

\subsection{UC3: Platooning Lead Services}
Platoons decrease the distances between the participating vehicles using wireless electronic coupling, thereby decreasing the aerodynamic drag and hence the fuel consumption \cite{SD14}. This capability would allow many vehicles to accelerate or brake simultaneously. It also allows for a closer headway between vehicles by eliminating distance needed for human reaction. \\

Platooning lead services can be considered to be a commercial VAS where pre-registered drivers (most preferable truck drivers) are allowed to act as lead vehicle for an arbitrary automotive platoon during the course of their journey. Hence, what we have are two sets of drivers---one acting as a lead vehicle and the others as followers. The rational for a registered lead vehicle driver is to have a qualified professional rather than an arbitrary person.

\subsection{UC4: Visible Light Communication}
VL Communication is generating a lot of interest today particularly owing to its ease of implementation due to the existence of infrastructure (40 billion light bulbs compared to 1.4 million base stations currently being deployed) and larger spectrum (10,000 times compared to radio spectrum). VLC in a primitive sense is already being used in vehicles in the form of active headlights and anti-glare rear view mirror where a photo diode is used to detect any vehicle approaching from front and dip/curve the headlights automatically. The same principle is also used to lower the rear-view mirror in case of headlight flashing by the vehicle in rear.

This concept can be extended to be used as a communication device. The head and tail lamps including traffic signal lights can be modulated with respect to amplitude/spectrum so as to also carry data. Such a modulation should be invisible to the naked eye and can only be detected by photodiodes.

Robust point to point light links can also be created between vehicles moving one after the other to transmit with low to high bitrates. VLC can also be used to detect pedestrians and warn them of an approaching vehicle beforehand. In this case, the mobile phones equipped with photodiodes can detect an approaching vehicle and warn the pedestrian. The mobile phone can also broadcast this information to street lights and other users nearby using the radio Device-to-Device (D2D) mode. In another implementation, the street lights equipped with photodiodes can also disseminate warnings about approaching vehicles to the pedestrians walking below it.

\subsection{UC5: Fuel-Optimal Assisted and Automated Driving}
Assisted and automated driving can considerably reduce the fuel consumption \cite{LGL14}. The concept relies on a fuel-optimal operation of the vehicle based on a vehicle model and environment information obtained from V2V and V2I communications. For example, static information like road grades and dynamic information like traffic light states can be used to control the vehicle speed in a fuel-optimal way. The simulation and evaluation platform provides a versatile tool for evaluating such concepts in a realistic setting. Particularly, the traffic effects in such concepts can be quantitatively investigated with a simulation and evaluation platform. Furthermore, communication aspects can be effectively studied.


\section{Cyber-Automotive Simulation and Evaluation Platform (CASEP)}
The fact and the possibility to realize a simulation without having to implement experiments in the real world leads to solutions that do not involve high investments, particularly for new or expensive infrastructure. Results we obtain from the simulation could be used to implement novel concepts to reduce traffic congestion improve road safety, and decrease fuel consumption.

The Cyber-Automotive Simulation and Evaluation Platform (CASEP) follows this purpose. Main modules are the
\begin{itemize}
	\item \emph{Communication Module} for simulating communication aspects
	\item \emph{Raytracing on Real Maps Module} for including geographic aspects
	\item \emph{Mobility Module} for generating vehicular and pedestrian traffic
	\item \emph{Vehicle Models Module} for computing the fuel consumption (detailed below)
	\item \emph{Driver Module} for generating driving profile
	\item \emph{Visualization Module} for visualizing results.
\end{itemize}
These modules can be combined in a modular and scalable manner. Particularly, tools from various domains (vehicle simulation, communication simulation, traffic simulation, visualization tools, etc.) can be combined in the platform in a flexible way.


%
%
%


Vehicle models are used in the platform for generating vehicle trajectories and computing the vehicle fuel consumption. In the following a model for conventional vehicles with internal combustion engine is described. The model can be easily extended for electric and hybrid vehicles.

The vehicle longitudinal dynamics is described by
\begin{align}
	 	F_\text{t} = (m_\text{v}+m_\text{r}) \dfrac{d v}{d t} + F_\text{r} + F_\text{a} + F_\text{g} + F_\text{hydr}
	 	\label{eqn:MechDynamic}
\end{align}
where $m_\text{v}$ is the vehicle mass and $v$ is the vehicle speed. During the translational movement several mechanical components rotate. Therefore the corresponding moments of inertia must be considered. These are summed up and then converted into an equivalent mass $m_\text{r}$. The aerodynamic friction is given by $F_\text{a}=c_1v^2$, where the constant $c_1$ depends on several factors, such as the density of the ambient air. The rolling resistance is approximated by $F_\text{r}=c_2\cos\alpha$, where $\alpha$ is the road angle and the constant $c_2$ is mainly dependent on road surface condition. The gravitational force is expressed as $F_\text{g} =m_\text{v}g\sin\alpha$, where the gravitational acceleration is denoted as $g$. The traction force is modeled by $F_{\text{t}}$, the hydraulic braking force is modeled as $F_\text{hydr}$.

For generating vehicle trajectories the vehicle longitudinal dynamics (\ref{eqn:MechDynamic}) is simulated and evaluated with the forces $F_{\text{t}}$ and $F_\text{hydr}$ as input and the vehicle speed $v$ as output ("forward approach"). For computing the vehicle fuel consumption the vehicle longitudinal dynamics (\ref{eqn:MechDynamic}) is simulated and evaluated with the vehicle speed $v$ as input and the forces $F_{\text{t}}$ and $F_\text{hydr}$ as output ("backward approach" \cite[pp. 39]{GS13}). The fuel consumption is then determined from the traction force $F_{\text{t}}$. Specifically, the engine torque is calculated from $T_\text{ICE} = \frac{F_\text{t}R_\text{w}}{\eta\gamma}$, where $R_\text{w}$ is the wheel radius, $\eta$ is the gearbox efficiency, and $\gamma$ is the gear ratio. The engine angular velocity is calculated from $\omega_\text{ICE}=\frac{v \; \gamma}{R_\text{w}}$. The instantaneous fuel consumption is then computed from the engine torque $T_\text{ICE}$ and engine angular velocity $\omega_\text{ICE}$ using a fuel consumption map. The overall fuel consumption finally results from integrating the instantaneous fuel consumption.



\section{Example Simulation}
The CASEP framework being a modular simulation platform uses various interfaces to external codes. Here we will provide an example where different tools are deployed in tandem under the framework. We study an urban V2V communication scenario under the public transport infrastructure.

\subsection{Mobility Traces with SUMO}

We have chosen the city of Kaiserslautern for this CASEP experiment. Practically any other test city can been chosen, the mandatory requirement being geographic information system (GIS) data comprising the road network graph and other related geo tags (stop geo location, point of interest (POI) names, etc.) which can be extracted e.g.~from the OpenStreetMap (OSM) \cite{OSM1} data set. Such city specific data is readily available in the OSM format under the open source license. The OSM data is used to simulate public transport mobility. We have only considered the bus routes. Given the GIS data the Simulation of Urban MObility (SUMO) \cite{Behrisch11sumo} tool generates the bus traces based on the timetable information. Altogether 56 unique bus routes were simulated. This trace data also included simulated traffic of 5000 private vehicles across the whole city region so that we can mimic traffic congestion. SUMO was employed to generate mobility traces for a complete working day.

\subsection{V2V Deployment Considerations}

\begin{figure}
\centering
\includegraphics[scale=.43]{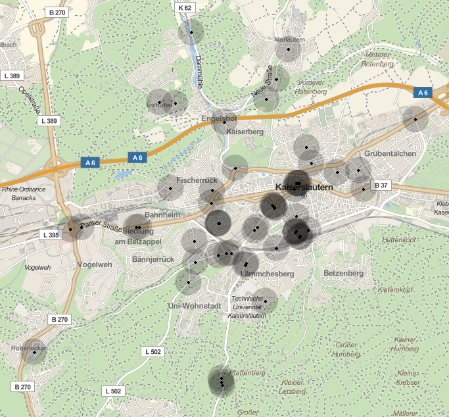}
\caption{The proximity inclusion disk for V2V communication initialization}
\label{fig:KL_rad}
\end{figure}

Given the public transport vehicle trace we came up with a V2V deployment scheme. The main goal was to study coordinated V2V connectivity among the bus constellation so that routing of message is possible from any source vehicle to another sink vehicle. In Fig.~\ref{fig:KL_rad}, the gray disks which depict the proximity inclusion threshold can be seen around each bus at a given time instant. 

\begin{figure}
    \centering
    \subfloat[Weighted adjacency matrix of the buses ]{{\includegraphics[scale=.45]{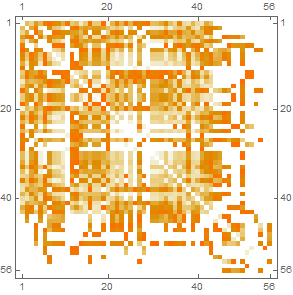} }}
    \qquad
    \subfloat[The community clustering showing long links]{{\includegraphics[scale=.20]{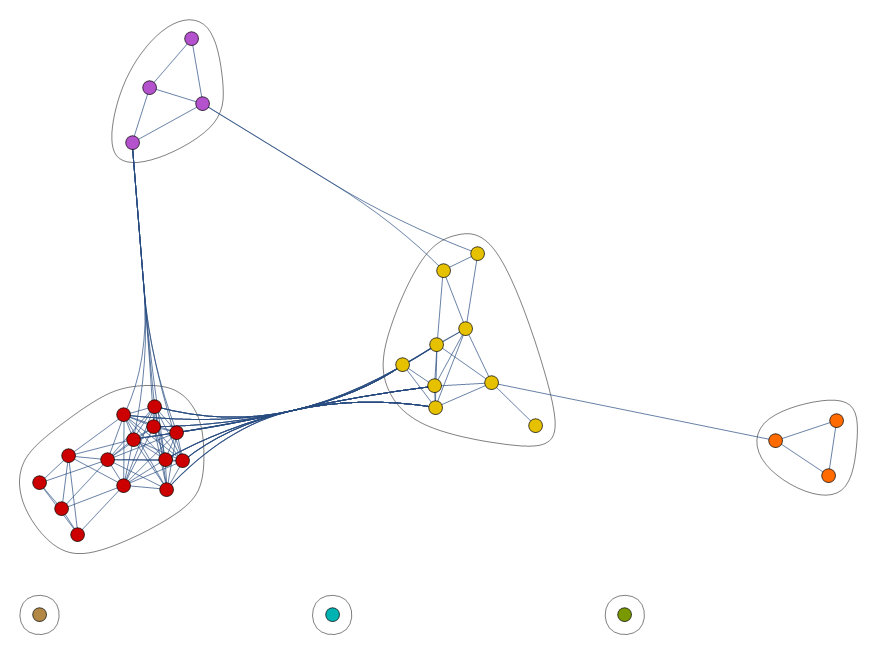} }}
    \caption{The underlying graph for V2V communication deployment}
    \label{fig:Graph}%
\end{figure}

\begin{figure}
\centering
\includegraphics[scale=.43]{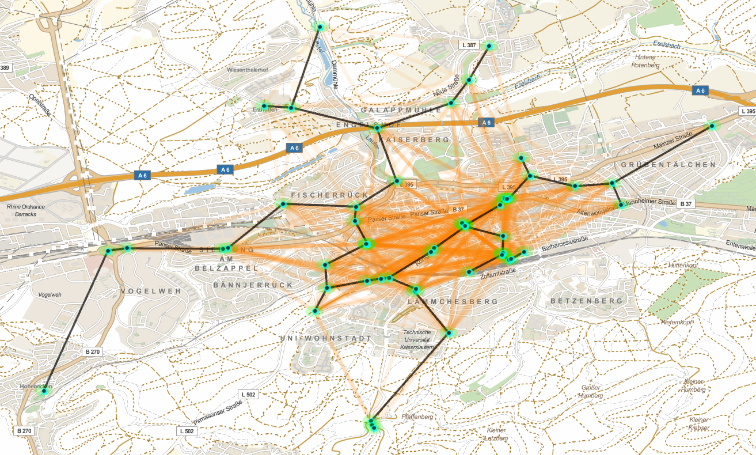}
\caption{Avoiding long radio links for V2V deployment}
\label{fig:Links}
\end{figure}

This threshold defines the distance limit within which a V2V connection can be established. However, to keep the graph connected some longer radio link initiation is permitted, though their occurrences are minimized. This process of avoiding longer V2V links generate locally clustered community graphs (Fig.~\ref{fig:Graph} (b)). The weighted adjacency matrix in Fig.~\ref{fig:Graph} (a) is nothing but the distance matrix for the vehicles at a certain time instant. One needs to find a spanning tree for this graph so that long links are avoided, keeping the underlying graph as much connected as possible. Also we put an upper limit of vertex connectivity of four. In Fig.~\ref{fig:Links} we show the possible V2V links in orange and the chosen link to form the spanning tree for the underlying graph in black. This connected network changes its topology as we loop through the SUMO mobility trace. The spanning tree is constructed in a time-dependent fashion and as seen in Fig.~\ref{fig:Graph} (b) some nodes/buses can become unreachable at times.


\begin{figure}
\centering
\includegraphics[scale=.2]{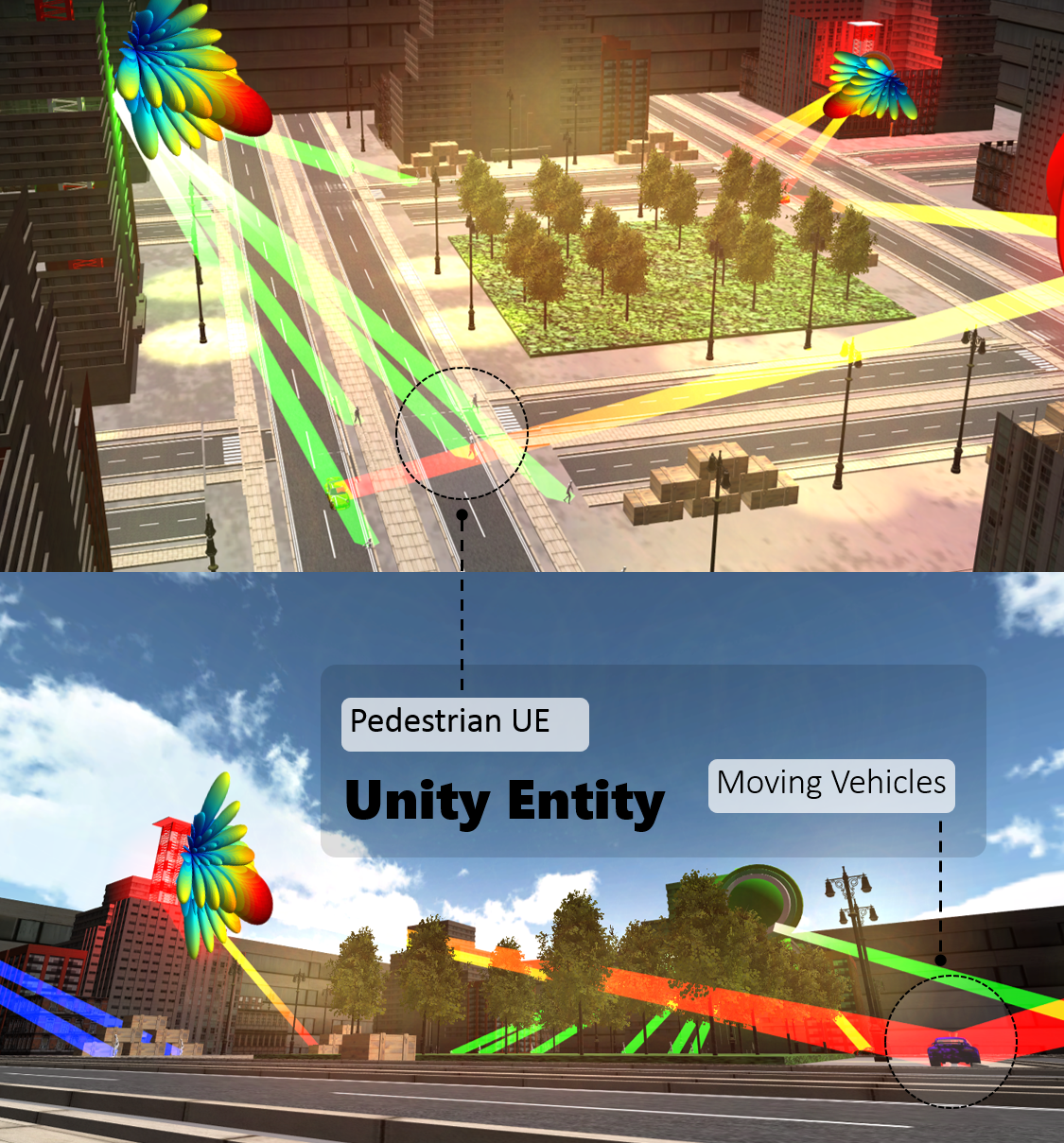}
\caption{Visualization example under the CASEP framework}
\label{fig:Visualization}
\end{figure}

\subsection{Interactive Unity3D Visualization}
We visualized the V2V network deployment along with the SUMO mobility traces using CASEP's Unity3D tool. The resulting visualization can be seen in Fig.~\ref{fig:Visualization}. Imaginary buildings were construed for visualization purposes and car traces as well as pedestrian movement were shown. The radio links between pedestrians and micro base stations can be seen. V2V links are also visualized in the same fashion. This was a conceptual validation of the CASEP visualization tool where hundreds of users were simulated as tradition mobile UE along with SUMO vehicle trace. We also tested the possibility of visualizing the waveform which is important for beam forming applications.
\section{Conclusions}
In this paper a novel cyber-automotive simulation and evaluation platform (CASEP) has been presented. The platform allows investigating the interactions between vehicles and the environment with high flexibility, modularity, and scalability. Particularly, tools from different disciplines, such as vehicle simulators, communication simulators, traffic simulators, and visualization frameworks, can be connected with the platform. Several use cases have been outlined, including investigations on peak power balancing, mobile base stations, platooning, visible light communications, and assisted and autonomous driving. As a detailed example a study on V2V communications has been presented.


\bibliographystyle{unsrt}
\bibliography{references}

\end{document}